\documentclass[runningheads]{llncs}
\usepackage[T1]{fontenc}
\usepackage{graphicx}
\begin{document}
\title{From Explanation to Diagnosis: Next Generation Interactive Video Coach with Misstep Awareness\thanks{Accepted at the 2026 AIED Workshop on Designing and Evaluating Next-Generation Learning Interfaces: Linking AI, HCI, and the Learning Sciences.}}
%
%
\author{Xiao Jin\and
Rahul K. Dass\and
Ashok K. Goel}
\authorrunning{X. Jin et al.}
%
\institute{Georgia Institute of Technology, Atlanta GA, USA 
\email{\{xjin96,rdass7,ag25\}@gatech.edu}}
\maketitle              
\begin{abstract}
Intelligent tutoring systems excel at generating explanations but rarely provide principled diagnosis of where and why a learner is wrong. We introduce a misstep-aware coaching capability for Ivy, a neurosymbolic AI coach, built on a two-model architecture that augments a Task-Method-Knowledge (TMK) model with a new Pedagogical Model (PM) in the context of an online graduate AI course at Georgia Tech. The PM makes instructor diagnostic knowledge explicit and machine-readable by encoding, for each quiz question and incorrect response, the learner's underlying belief(a brief statement of the incorrect idea or missing knowledge), a TMK locus(the source of the misunderstanding), a misconception type and targeted scaffolding derived from the instructor's Q\&A key. Using quiz questions from the course, we demonstrate a proof-of-concept pipeline that detects and classifies learner errors and generates diagnosis-grounded scaffolding, moving Ivy beyond knowledge retrieval toward diagnostic misstep awareness, and enabling more precise, actionable feedback that supports conceptual change and advances adaptive learning systems in AI in education and the learning sciences.

\end{abstract}
\section{Introduction}
The rapid growth of online education has fundamentally changed how adult learners access higher education and workforce training~\cite{means2020suddenly,castro2021literature}. Unlike in-person classrooms, online learning is often a more passive experience\cite{Chi02102014}. Students need real-time, bidirectional interaction so instructors can observe their reasoning process, identify errors as they form, and intervene before misconceptions solidify. Intelligent tutoring systems (ITSs) have demonstrated the potential to improve student performance through timely and individualized feedback~\cite{mousavinasab2021intelligent}. However, they remain limited in their ability to infer the underlying causes of learner errors. This diagnostic gap constrains the effectiveness of feedback, as meaningful instructional support depends not only on explaining what went wrong, but also on identifying why the error occurred~\cite{kuhail2023interacting}.

The Interactive Video or Ivy~\cite{dass2025ivy} is an AI coach embedded alongside instructional videos deployed in a graduate hybrid AI course in Georgia Tech's Master of Computer Science program. It combines a large language model with a structured knowledge representation called a Task-Method-Knowledge (TMK) model\cite{murdock2008meta,goel2017gaia}. Ivy's current architecture is a four-stage constrained generation pipeline that classifies learner queries, retrieves structured TMK entries, synthesizes constrained explanations, and optimizes coherence~\cite{lum2025designing}. It effectively explains what a correct step looks like and why it serves the goal. This paper describes the design and proof-of-concept instantiation of a two-model architecture that enables Ivy to move from explanation to misstep-aware coaching.

\vspace{1em}

\noindent\textbf{RQ:} Can a pedagogical model(PM) fill the diagnostic gap in intelligent tutoring systems by classifying learner missteps?

\textbf{RH:} PM-grounded Ivy will outperform TMK-only Ivy on feedback targeting, actionability, transferability, and scaffolding appropriateness, because error classification and underlying belief identification enable feedback that address the learner's specific misstep rather than the skill topic in general.

\section{Misconception Detection in Educational AI}

Recent work has begun applying LLMs to misconception detection in specific domains. Fathi found that integrating LLM within a inquiry-based framework significantly improved first-year engineering students’ conceptual understanding of thermodynamics and reduced misconceptions~\cite{el2025integrating}. Dimeren demonstrates that transformer-based and clustering methods can automatically detect and classify students’ physics misconceptions with near-expert accuracy, showing strong potential for AI to replace time-intensive human diagnosis of learner misunderstandings~\cite{demirezen2023new}. Kokver discovered that AI-based natural language processing models can effectively identify teacher candidates’ misconceptions about the greenhouse effect, offering a scalable alternative to expert human evaluation for diagnosing conceptual understanding~\cite{kokver2025artificial}. 

While existing approaches in AI in education focus on content delivery or surface-level error detection, they often fail to identify learners' misconceptions. Our approach offers a novel framework to address the learners' specific misconceptions, enabling more targeted feedback and supporting deeper, sustained learning.

\section{The Two-Model Architecture}
\subsection{TMK-only Ivy Architecture}
Ivy's architecture operates through four sequential stages~\cite{dass2025ivy,dass2025improving}. First, a question answerability module filters inappropriate or out-of-scope queries. Second, a knowledge retrieval module retrieves the top-k most relevant documents from a given TMK model. Third, a response generation module iteratively synthesizes an answer by incorporating information across the retrieved documents. Finally, a response optimizer refines the output for clarity and conciseness, tailoring verbosity to the nature of the query. This pipeline, powered by GPT-5 nano model via LangChain, allows Ivy to generate structured, accurate, and contextually aligned explanations that consistently outperform standard GPT and RAG-based baselines in system-level developer evaluations\cite{dass2025ivy,dass2025improving}.

A TMK model for a skill encodes three interlocking structures~\cite{murdock2008meta,goel2017gaia,chandrasekaran1992task}. Task specifies the goal, its success criteria, and the conditions under which it is pursued. Method encodes a finite-state machine whose states represent world configurations and whose transitions represent executable steps with preconditions and postconditions. Knowledge defines the domain concepts. TMK encodes only the correct procedure with no representation of incorrect beliefs, and therefore cannot differentiate procedural errors.

\subsection{Pedagogical Model: Schema and Structure}
ITSs that rely solely on knowledge retrieval risk generating responses that are factually correct yet pedagogically ineffective, as they lack an explicit model of why a student has erred. The pedagogical model (PM) is introduced to address this gap by equipping Ivy with a structured diagnostic layer that moves beyond answer retrieval toward reasoned, misconception-aware tutoring. The PM is a structured JSON artifact authored by instructors for each quiz questions. Each PM record contains the following fields showing in Table 1.

\begin{table}[h!]
\centering
\setlength{\tabcolsep}{6pt}
\caption{Fields included in the PM record schema.}
\label{tab:pm_record_schema}
\begin{tabular}{|l|p{9cm}|}
\hline
\textbf{Field} & \textbf{Description} \\
\hline
question\_id & Unique identifier linking the PM record to the assessment item. \\
\hline
question\_text & The full text of the question as presented to the learner. \\
\hline
options & A list of answer choices, each associated with a correctness flag. \\
\hline
error\_type & One of the four taxonomy categories defined in Section 3.3. \\
\hline
tmk\_locus & The specific TMK entry at which the learner’s error is localized. \\
\hline
underlying\_belief & Statement of the incorrect belief or missing knowledge. \\
\hline
scaffolding & The recommended feedback form and content. \\
\hline
\end{tabular}
\end{table}

When a learner selects an incorrect answer, Ivy looks up the corresponding PM record, reads the \texttt{error\_type} and \texttt{tmk\_locus}, and selects the scaffolding entry as the seed for its constrained synthesis stage. The LLM then generates a response grounded in both the retrieved TMK entries and the PM scaffolding template, ensuring that the explanation is both procedurally faithful and targeted to the diagnosed misconception. More details about PM construction is described in section 4.

\subsection{The Four-Category Misstep Taxonomy}
The taxonomy draws on Norman's slip-mistake distinction~\cite{norman2013design}, Reason's error classification~\cite{reason1990human}, Chi and Roscoe's conceptual-change framework~\cite{Chi2002}, and Newell and Simon's task decomposition theory~\cite{newell1972human}. While these frameworks were developed in general cognitive contexts, we adapt and synthesize them specifically for AI education. Our taxonomy extends prior work by mapping these errors to specific TMK signal, enabling targeted feedback and instructional intervention. Four categories are defined in Table 2.

\begin{table*}[h!]
\centering
\setlength{\tabcolsep}{4pt}
\renewcommand{\arraystretch}{1}
\caption{Misstep taxonomy with definitions, TMK signals, and feedback focus.}
\label{tab:misstep_taxonomy}
\begin{tabular}{|p{1.7cm}|p{3.2cm}|p{3.2cm}|p{3.2cm}|}
\hline
\textbf{Error Type} & \textbf{Definition} & \textbf{TMK Signal} & \textbf{Feedback Focus} \\
\hline

Execution Slip &
Local notation or syntax error with intact plan and reasoning. &
Local mismatch; preconditions and goal alignment remain valid. &
Corrective cue identifying the specific fix, preserving the plan. \\
\hline

Conceptual Gap &
A concept or mechanism is missing &
Missing knowledge, cannot explain preconditions &
Explain missing concept or constraint; understanding check. \\
\hline

Causal Misunderstanding &
Violates a required condition or the claimed effect. &
Violated precondition or incorrect postcondition in the method transition. &
Clarification of the violated causal link and its consequence. \\
\hline

Teleological Error &
Jumps to an
inapplicable technique;
omits required subgoal. &
Steps or subgoals are misordered, preventing progress toward the goal. &
Restatement of the goal and guidance toward the correct subgoal. \\
\hline
\end{tabular}
\end{table*}

\vspace{-2em}
\subsection{Pairing TMK and PM: The Diagnosis Pipeline}
When a learner submits an answer or question, Ivy executes a five-step diagnosis pipeline. First, it classifies the query scope same as the existing TMK only Ivy. Second, if a quiz submission is detected, the system looks up the PM record. If the similarity score is below a threshold (0.5), indicating that no matching PM was found, the process falls back to the TMK-only Ivy pipeline for TMK retrieval. Otherwise, it proceeds to the third step. Third, it retrieves the relevant TMK entries specified by the \texttt{tmk\_locus}. Fourth, it synthesizes a scaffolding response constrained by both the PM scaffolding template and the retrieved TMK entries. Fifth step is the same as the existing one. Finally, the system logs the identified error type, the associated TMK locus, and the generated response. The diagram (Figure 1) summarizes how the pedagogical model layer integrates with Ivy's existing four-stage pipeline at runtime. Shaded boxes are new components introduced.

\begin{figure}[h]
    \caption{Ivy Pedagogical Model Integration Pipeline}
    \label{fig:my_label}
    \centering
    \includegraphics[width=0.7\textwidth]{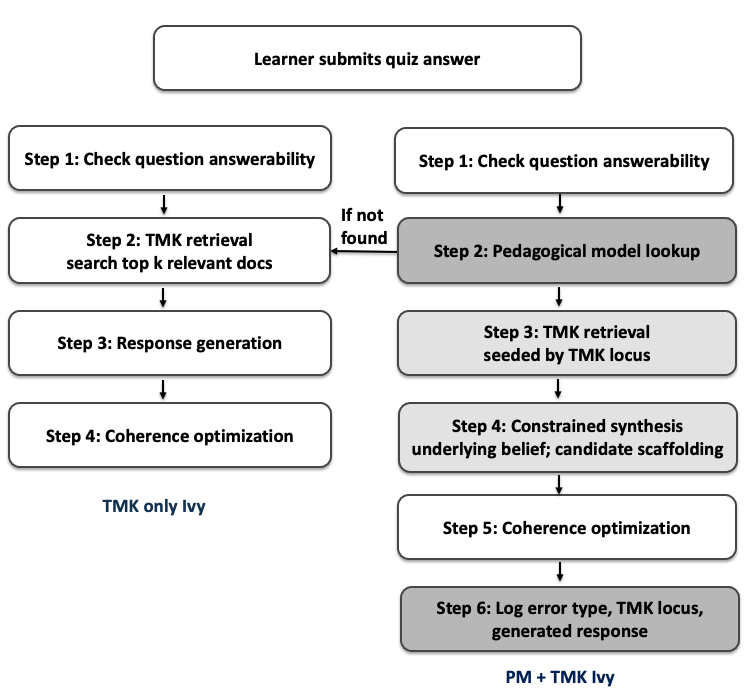}   
\end{figure}

\vspace{-0.5em}
\section{Proof-of-Concept Instantiation}
\subsection{Source Material}
PMs were constructed from quiz questions in the Spring 2026 offering of CS 7637: Knowledge-Based AI(KBAI) at Georgia Institute of Technology. The source material comprised: (1) the full text of each question and its answer choices; (2) the expert authored Q\&A key specifying the correct answer and, for each incorrect answer, a brief rationale for why a learner might select it; and (3) the TMK models previously authored for the three targeted skills. In total, 9 questions spanning three skill areas were processed.

Three skill areas were instantiated, yielding 9 PM records and 22 incorrect answer pairs: MetaReasoning (1 question, 3 incorrect answers), Production Systems (5 questions, 13 incorrect answers), Scripts (3 questions, 6 incorrect answers). 

\subsection{PM Construction Process}
Each PM file captures a diagnostic profile of a specific learning failure, including an error type, a TMK locus identifying where corresponding knowledge representation occurs, the underlying false belief driving the error, and scaffolding strategies for remediation. These files are stored in a dedicated repository and indexed at system initialization into a FAISS vector store, where each PM document is embedded in a OpenAI's text embedding model. This one-time indexing step transforms the expert-curated misconception library into a retrievable semantic index, enabling efficient nearest-neighbor lookup at inference time without repeated embedding overhead.

Upon receiving a student query, the system computes its semantic embedding and performs a similarity search against the PM index, retrieving the single most proximate misconception archetype subject to a minimum relevance threshold. If a candidate PM clears this threshold, it is passed alongside the student's question to a large language model acting as a zero-shot classifier, which determines whether the retrieved archetype genuinely characterizes the student's error and populates the remaining diagnostic fields. A confirmed match causes the system to bypass the TMK vector store entirely, instead injecting the structured pedagogical context directly into the generation prompt, thereby grounding the tutoring response in a theoretically motivated model of the student's specific misconception rather than generic domain knowledge retrieval.

\section{Preliminary Evaluation}
\subsection{Error Classification Analysis}
Across all 22 incorrect answers, the distribution of error types was: causal misunderstanding 45\% (n=10), teleological error 14\% (n=3), conceptual gap 23\% (n=5), and execution slip 18\% (n=4). To obtain a preliminary estimate of PM-guided diagnosis coverage, we constructed an evaluation task using the 9 questions across three skills from the KBAI course quizzes. For each question, all candidate answer options were systematically enumerated and submitted independently to the Ivy PM-guided diagnosis pipeline. Each submission simulated a student selection, allowing exhaustive evaluation across all possible response trajectories within the domain.

The pipeline matched each response against the pedagogical model using similarity search, retrieving the most semantically relevant PM record and generating a predicted error-type diagnosis for each response. Across all question-answer pairs, the pipeline achieved 100\% PM retrieval coverage. Every submitted response was successfully matched to a PM record and assigned a corresponding error-type label. No response fell below the similarity threshold or failed to retrieve a valid PM entry.

\subsection{Scaffolding Quality Comparison}
To assess whether PM adds diagnostic value beyond Ivy's existing TMK-only pipeline, we conducted scaffolding quality comparison study. For each of the 38 answer options from 9 questions, we generated two candidate feedback responses: one using Ivy’s existing TMK-only pipeline and one using the PM-augmented pipeline. Raters evaluated each response on five dimensions using a 3-point scale: 0 (incorrect), 1 (partially correct), and 2 (correct). The dimensions as follows:

\begin{itemize}
    \item \textbf{Accuracy:} Whether the feedback is correct.
    
    \item \textbf{Targeting:} Whether the feedback addresses the specific misconception.
    
    \item \textbf{Actionability:} Whether the feedback provides a clear path to attempt.
    
    \item \textbf{Transferability:} Does the feedback help the learner recognize when the skill applies in a new context.
    
    \item \textbf{Scaffolding Appropriateness:} Whether the form of feedback matches the nature of the learner's error.
\end{itemize}
Table 3 reports mean scores (0–2 scale) and paired comparison p-values across five feedback quality dimensions for TMK-only and PM-augmented Ivy responses. PM-grounded responses outperformed TMK-only responses on all five dimensions, with statistically significant gains on four of the five.

\begin{table}[h!]
\centering
\small
\setlength{\tabcolsep}{4pt}
\caption{Comparison of average feedback quality(n = 38).}
\begin{tabular}{lccccc}
\hline
 & Accuracy & Targeting &
 \begin{tabular}{c}Action-\\ability\end{tabular} &
 \begin{tabular}{c}Transfer-\\ability\end{tabular} &
 \begin{tabular}{c}Scaffolding\\Appropriateness\end{tabular} \\
\hline
TMK only & 1.42 & 0.95 & 0.47 & 0.55 & 0.71 \\
PM + TMK & 1.58 & 1.50 & 1.32 & 1.32 & 1.42 \\
p value & 0.2712 & 0.0011 & 1.22E-06 & 1.33E-05 & 3.91E-05 \\
\hline
\end{tabular}
\end{table}

The most significant improvements appeared on actionability (M = 1.32 vs. M = 0.47, p < .001) and transferability (M = 1.32 vs. M = 0.55, p < .001). This pattern is interpretable: without a PM record specifying the underlying incorrect belief and a recovery path, TMK-only responses tend to re-explain the correct procedure but leave the learner without a concrete next step or a sense of when the skill applies in a new context. Scaffolding appropriateness showed a similarly strong effect (M = 1.42 vs. M = 0.71, p < .001), indicating that PM-grounded responses more reliably matched the feedback form to the diagnosed error type, consistent with the theoretical motivation for the taxonomy.

Targeting also improved significantly (M = 1.50 vs. M = 0.95, p = .001), confirming that PM-grounded responses more consistently addressed the learner's specific error rather than the general skill topic. Accuracy was the one dimension that did not reach statistical significance (M = 1.58 vs. M = 1.42, p = .271), suggesting that both pipelines produce the correct characterizations of the skill. The TMK model alone is sufficient for factual correctness, although that correctness alone does not translate into targeted, actionable, or transferable feedback without the diagnostic layer the PM provides.

These results indicate that the diagnostic gap in TMK-only coaching is not a gap in factual accuracy but in pedagogical specificity. Knowing the correct procedure is not sufficient to tell a learner why their particular move was wrong, what to do next, or how to recognize a similar situation in the future. The PM layer addresses precisely these deficiencies

\section{Discussion and Limitations}
The strong effects on actionability and transferability are particularly notable. TMK-only responses score (M = 0.47 and M = 0.55 respectively), suggest that without the underlying belief and scaffolding template, Ivy defaults to re-explaining the correct procedure. The response that may confirm what the learner should have done without helping them understand why they did not, or how to proceed. PM-grounded responses raised both dimensions to M = 1.32, a shift that raters described qualitatively as the difference between being told the answer and being shown a path forward. The improvement in scaffolding appropriateness (M = 0.71 to M = 1.42) further suggests that the four-category taxonomy provides appropriate feedback forms in a way that TMK signals alone cannot reliably support.

Nonetheless, several limitations bound the interpretation of these findings. The evaluation sample is small. Twenty-two responses spanning across three skills provide sufficient basis for a proof-of-concept but not for generalizable claims about PM performance across skill domains, question types, or learner populations. In addition, this study assesses only response quality as judged by expert evaluators. It does not measure whether PM-grounded scaffolding actually helps students correct their errors, reduces error recurrence within a session, improves post-test performance, or raises self-efficacy. 

\section{Conclusion and Next Steps}
This paper has presented the design and proof-of-concept instantiation of a pedagogical model that extends Ivy's TMK-grounded coaching architecture from explanation generation to misstep-aware diagnosis. The PM schema enables Ivy to classify learner errors into one of four theoretically grounded categories, localize the error within the skill's goal hierarchy, infer the underlying incorrect belief, and deliver targeted scaffolding response.

We are currently working on scaling this next generation AI coaching agent from proof-of-concept to a fully deployed and validated system. We are scaling it to the full course and ultimately across multiple courses. In addition, we will expand the PM framework beyond multiple-choice questions to the full range of question types. Open-ended submissions require a different diagnostic approach in which Ivy must infer the error type from the learner's free-form reasoning rather than look it up in a table. Extending the PM schema to encode error-type signatures combined with LLM-assisted classification constrained by TMK traces, is the natural next step. Furthermore, we would implement learner-model update across sessions. The current architecture diagnoses each submission independently, with no memory of the learner's prior error patterns. A longitudinal learner model would track error-type recurrence for each individual. Learners who repeatedly exhibit the same error type would receive progressively more explicit intervention. Implementing this requires a persistent learner state representation linked to the PM taxonomy, session-level aggregation of error-type counts, and a scaffolding policy that conditions on both the current diagnosis and the learner's accumulated error history. These extensions would help evolve Ivy into a dynamic longitudinal coaching system that supports learners throughout their learning journey.

\begin{credits}
\subsubsection{\ackname} This research has been supported by US NSF Grants \#2115232 and \#2247790 to the National AI Institute for Adult Learning and Online Education (aialoe.org). 
We thank members of the Ivy project team at Georgia Tech's
Design Intelligence Laboratory for their contributions to this work.

\end{credits}

\bibliographystyle{splncs04}
\bibliography{references}

\end{document}